\documentclass[aps,pra,reprint]{revtex4-2}

\usepackage{amsmath}
\usepackage{amssymb}
\usepackage{graphicx}
\usepackage{hyperref}
\usepackage{blindtext}

\begin{abstract}
The strong electric fields from tightly-focused and ultrashort laser beams have always been discussed as a way to accelerate charged particles without any need for a medium or external cavity. Radially-polarized light is one way to do this, motivated by the emergence of longitudinal electrical fields with tight focusing. However, the laser pulse will generally quickly overtake the electrons under its influence, and every-other half-cycle will decelerate the electrons in effect partially reversing the acceleration. In this work we present the effect of optical aberrations, primarily spherical aberration, and how despite their purely spatial nature they can significantly optimize the net acceleration, and advantageously allow for longer pulses to drive this optical field-based process. We discuss the optical physics responsible for this increase in performance and find optimal aberration profiles using a stochastic algorithm.
\end{abstract}
\begin{document}

\title{Optimization of vacuum acceleration with radially polarized laser beams having phase aberrations}
\author{Spencer W. Jolly}
\email{spencer.jolly@ulb.be}
\affiliation{Service OPERA-Photonique, Université libre de Bruxelles (ULB), Brussels, Belgium}
\date{\today}
\maketitle

\section{Introduction}

The acceleration of electrons to relativistic energies in vacuum is possible with a tightly focused, high power radially polarized laser beam (RPLB). The solution of the wave equation in focus emits a purely longitudinal field on axis~\cite{salamin06}, enforced by the symmetry of the tightly converging radial fields. This provides the motivation for using these beams to accelerate particles to relativistic energies in a true vacuum~\cite{varin05,salamin07}, sometimes referred to as RP-VLA. We numerically simulate the effects on such acceleration with a fundamental RPLB having spherical aberration, and other phase aberrations, showing how very simple changes can have unexpected and significant effects on this ultrafast and highly nonlinear process.

The most general case of vacuum laser acceleration by an RPLB is that where a test electron begins from rest (or with a moderate initial energy) and an RPLB focuses and overtakes the particle. The longitudinal field in the focal region, when the laser power is large enough, imparts a net kinetic energy on the particle after overtaking~\cite{wong10}. Studies were also done including non-paraxial terms~\cite{marceau12}, off-axis fields~\cite{marceau13-2} and even more complex interactions~\cite{sell14,varin16,wong17-3} showing that low energy-spread and collimation are indeed possible in a bunch of electrons with finite charge and size. Experimental results of this nature have been achieved in a low density neutral gas achieving 23\,keV energies~\cite{payeur12}, and in a true vacuum, accelerating some electrons in a bunch from 40\,keV up to a maximum of 52\,keV~\cite{carbajo16}. Recent experiments have shown MeV level energies either from plasma mirror injection~\cite{zaim17,zaim20} with high energy lasers or via optimization of the ionization process that occurs within the accelerating pulse itself~\cite{powell24} with a lower energy laser.

Beyond the use of a single Fourier-limited pulse-beam it is possible to optimize the interaction with more complex or structured pulses (structured beyond the radial polarization). A study was done in the general case using a beam composed of components of two colors with independent CEP~\cite{wong11-1}. It was shown that adding longitudinal chromatism (LC) in the focus, equivalent to pulse-front curvature (PFC) for an input transform limited pulse, can optimize the net acceleration in combination with various levels of chirp~\cite{jolly19-1}. It has also been shown that the inclusion of frequency-varying beam parameters, i.e. chromatic amplitude terms, can also optimize the acceleration when the pulse duration becomes few-cycle~\cite{jolly20-2}. Finally, it has also been shown that changing the illumination profile of the focusing lens from the standard lowest-order RPLB will change the evolution of the Gouy phase~\cite{pelchat-voyer20,pelchat-voyer21}, which can also have significant effect on the acceleration process~\cite{pelchat-voyer22}. In this work we will show specifically how phase aberrations can have a different and more significant effect than those structurations made previously, and also approach this ultrafast and highly nonlinear problem with a stochastic optimization algorithm to find the optimal aberrations and to learn more about the physical process.

\section{Background}

The ideal lowest-order radially-polarized illumination at the focusing lens of focal length $f$ has an amplitude profile of $A(r)\propto (r/w_i)\exp(-(r/w_i)^2)$, where $w_i$ is the initial beam waist before focusing. If there are no phase aberrations this results in a strong longitudinal field at the focus, which at $r=0$ is known analytically to be

\begin{align}
\label{eq:E_z_an}
E_z^{(\textrm{an})}(z,t)=\sqrt{\frac{8P}{\pi\varepsilon_0 c}}\frac{e^{i(\omega_0(t-z/c)+\Phi_0)}e^{-(t-z/c)^2/\tau_0^2}}{z_R\left(1-\frac{iz}{z_R}\right)^2},
\end{align}

\noindent without any non-paraxial corrections, where $\Phi_0$ is the CEP and $P$ is the pulse power. In all of the following simulations we use pulses that have Gaussian spatial and temporal/spectral profiles, with characteristic widths $w_0$ and $\tau_0$ respectively (spectral width $\Delta\omega=2/\tau_0$), at a central wavelength of $\lambda_0=800$\,nm ($\omega_0=2\pi c/\lambda_0$) where the Rayleigh range is $z_R=\omega_0 w_0^2/2c=w_0f/w_i$. The relatively straightforward scaling to other central wavelengths is discussed in past work~\cite{wong10}, where the choice for this work is based on the most prevalent high-power short-pulse lasers.

\begin{figure}[htb]
	\centering
	\includegraphics[width=86mm]{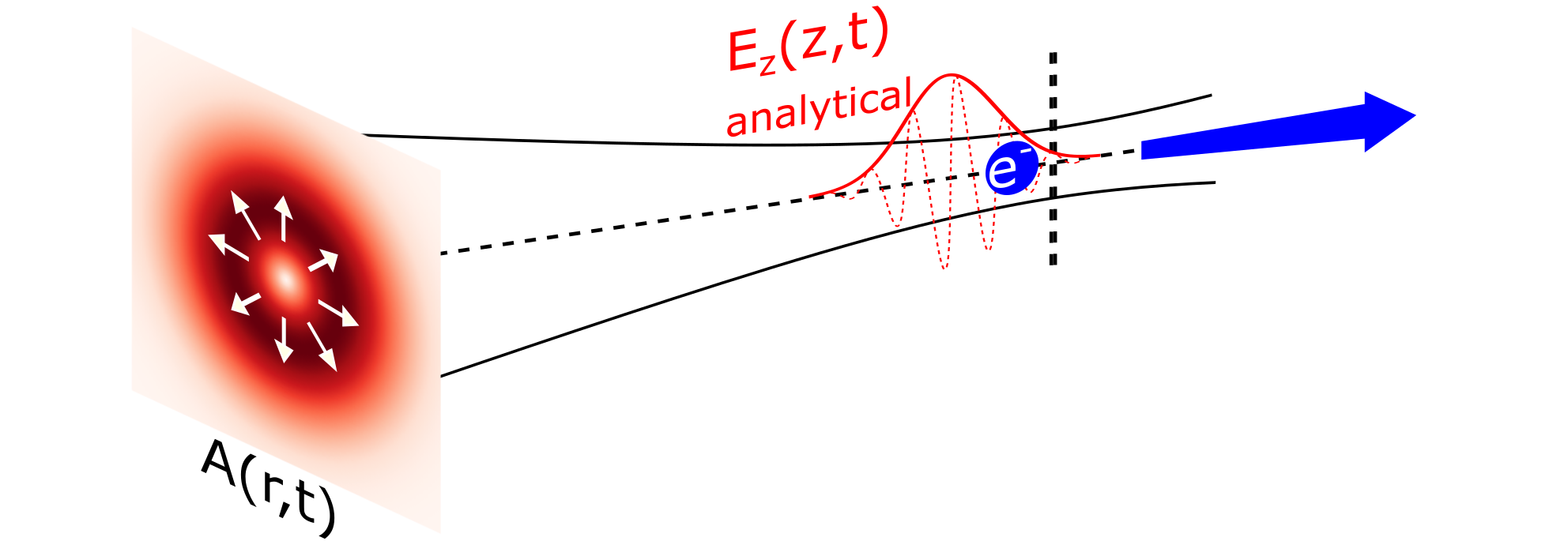}
	\includegraphics[width=86mm]{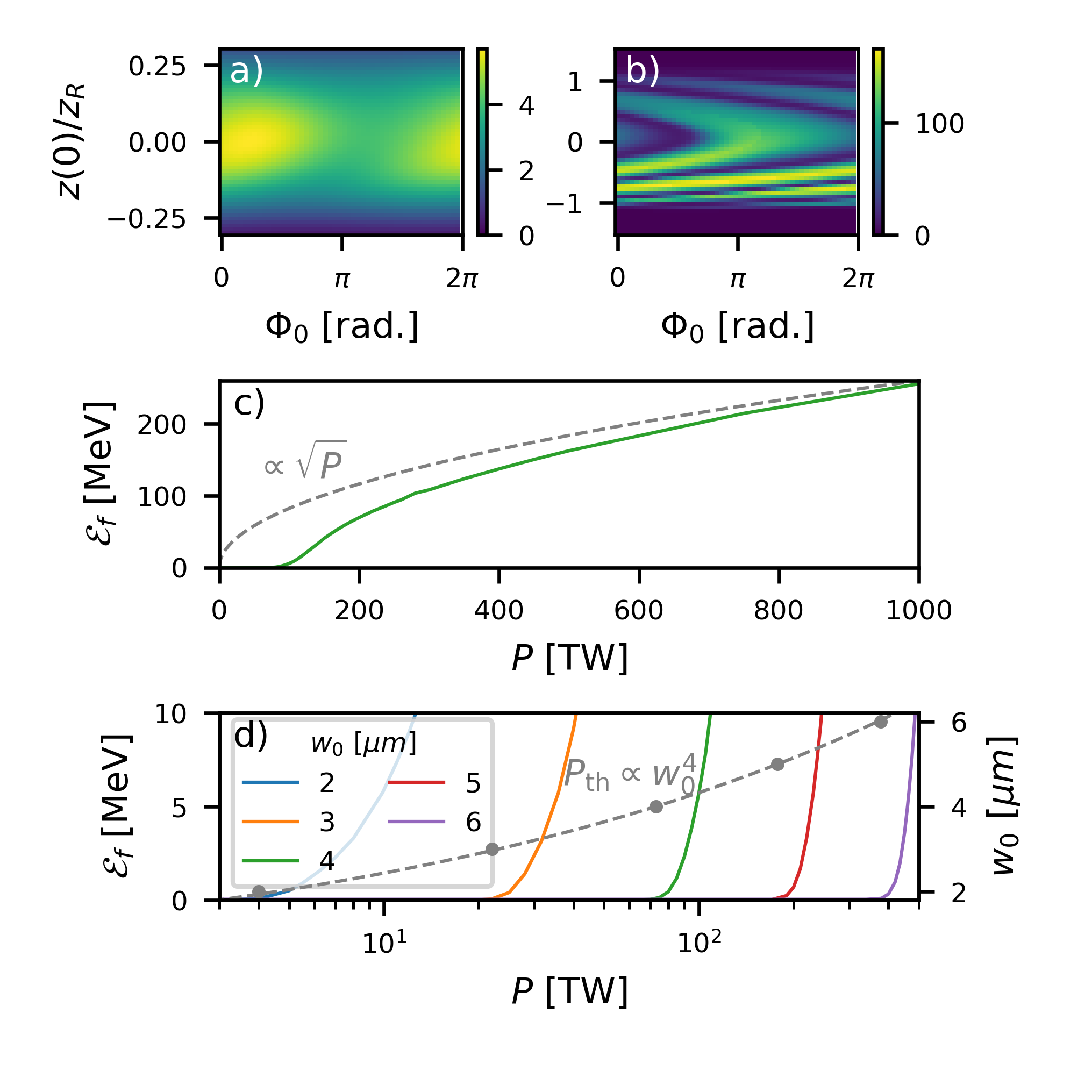}
	\caption{Summary of standard vacuum acceleration results without any spatial phase, sketched schematically (top). Acceleration with a 100\,TW pulse (a) to MeV energies shows weak periodicity with the CEP but the optimum is still roughly at $z(0)=0$ while with a 500\,TW pulse (b) the optimization is more complex with CEP and is at a significant negative $z(0)$. The scaling with pulse power (c) shows a threshold and eventual scaling with $\sqrt{P}$, while the threshold scales with $w_0^4$ (d). All plots shown are with $\tau_0=10$\,fs, while (a)--(c) have a waist of 4\,$\mu$m.}
	\label{fig:background}
\end{figure}

Detailed studies have been done showing especially that with optimization of both the initial position of the test particle relative to the waist plane $z(0)$ and the carrier-offset phase (CEP) of the driving laser $\Phi_0$---shown in Figure~\ref{fig:background}(a)--(b) for two laser powers---the final kinetic energy $\mathcal{E}_f$ of the particle is always higher with decreasing focused spot size (i.e. with increasing peak electric field)~\cite{wong10}. The optimal initial position becomes more negative and the behavior over CEP become much more strongly peaked when the peak field (and therefore the final energy) becomes larger (Fig.~\ref{fig:background}(b)). In this sense, when we are presenting the optimized final energy of a single electron for the rest of this work it represents the maximum energy attainable in what would usually be an ensemble of electrons that are experiencing the laser field as it propagates, and highly relevant for lasers that are CEP stabilized. Other scaling laws for the threshold and limitation of this mechanism were developed early-on~\cite{fortin10, wong11-2}, showing that the final electron energy scales with $\sqrt{P}$ far above the threshold (Fig.~\ref{fig:background}(c)) and the threshold itself scales with $w_0^4$ (Fig.~\ref{fig:background}(d)).

\begin{figure}[htb]
	\centering
	\includegraphics[width=86mm]{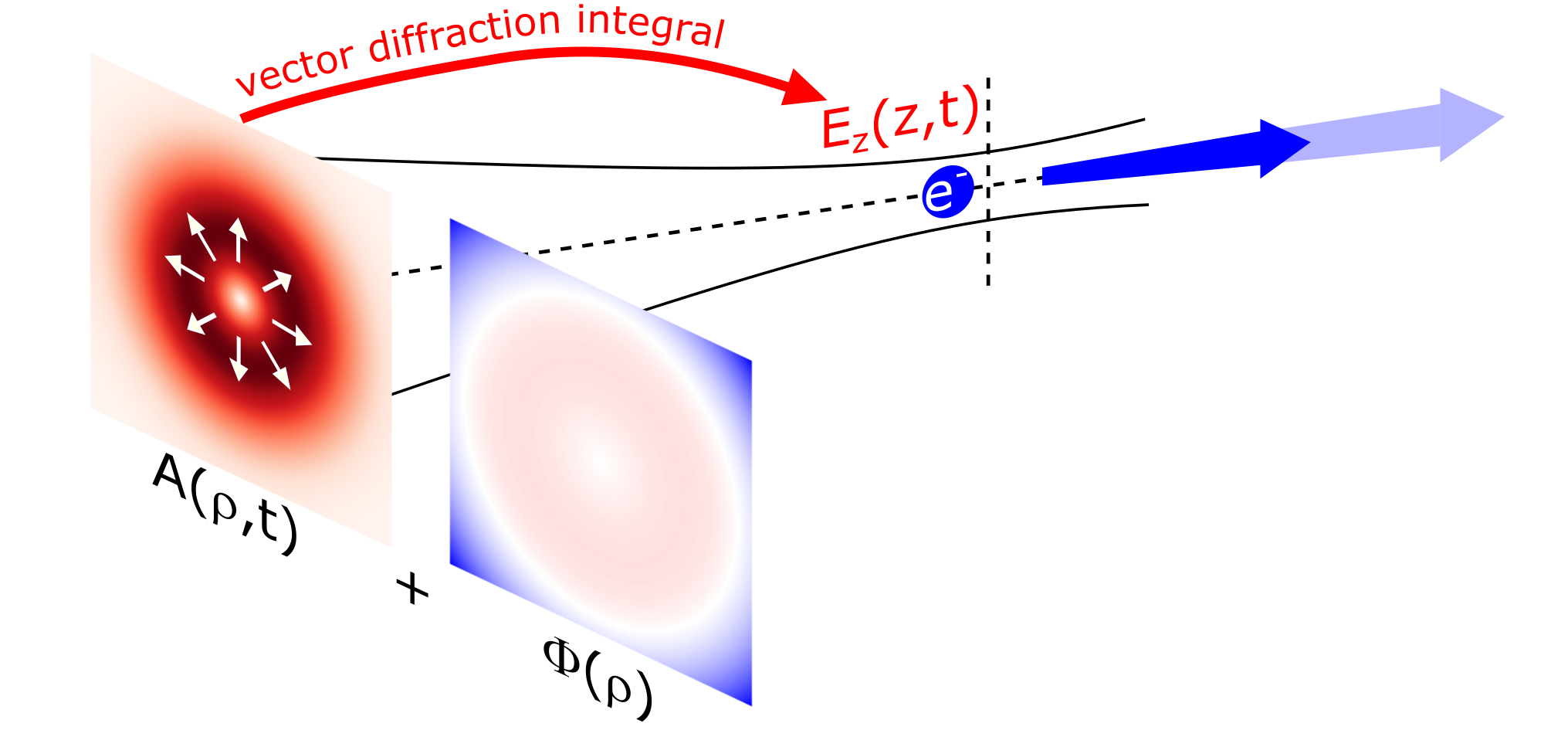}
	\includegraphics[width=86mm]{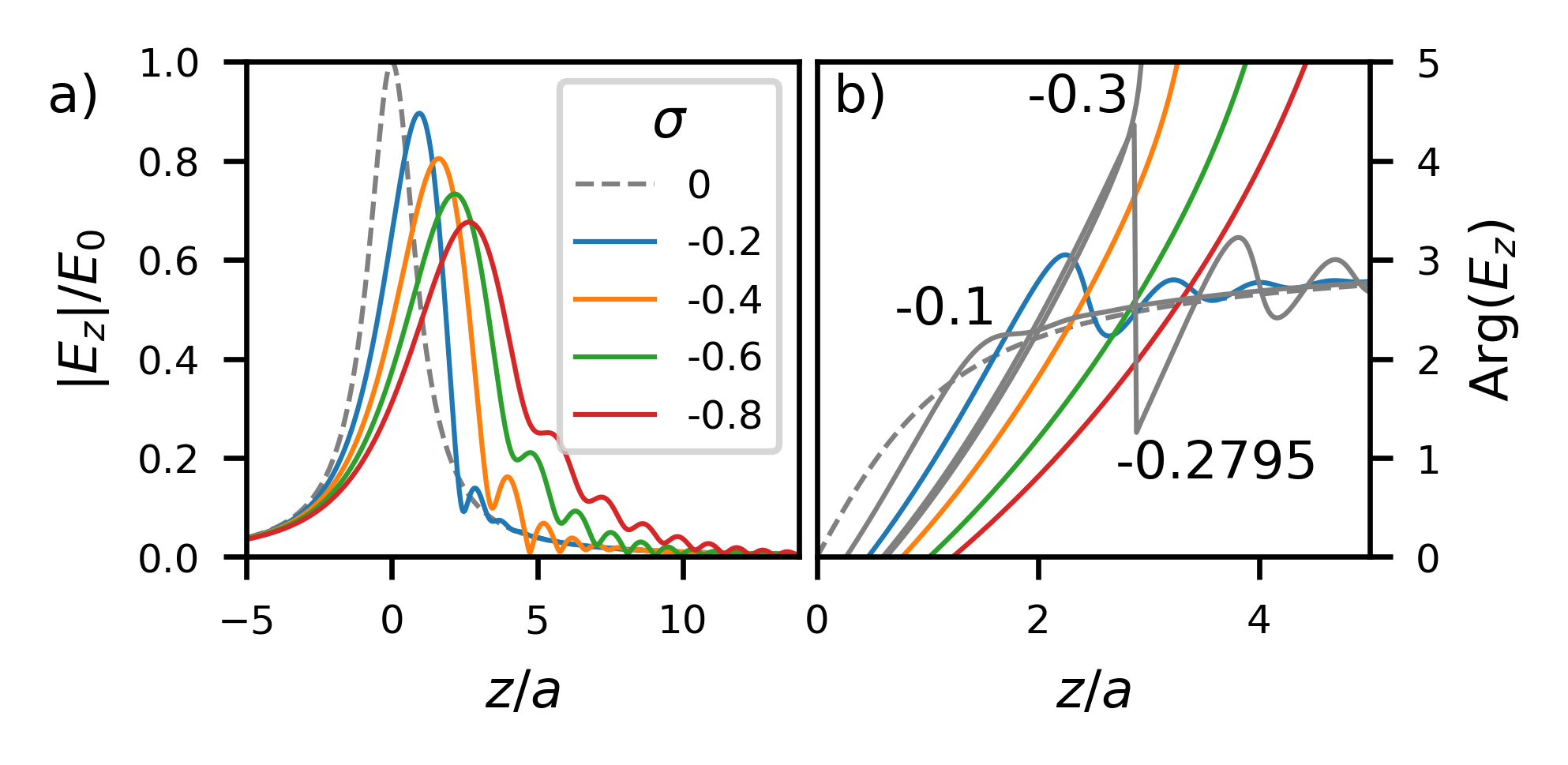}
	\caption{Summary of the scenario with an additional spatial phase, sketched schematically (top). The longitudinal field on-axis $E_z(r=0)$ (a) and the phase (b) are shown for a number of different amounts of spherical aberration $\sigma$ with $k_0a=125$, with additional curves manually noted in the case of the phase in (b).}
	\label{fig:spherical_field}
\end{figure}

Contrary to the work done on tuning the Gouy phase by changing the amplitude profile before focusing, which can significantly effect the acceleration~\cite{pelchat-voyer22}, in this manuscript we will study the phase aberrations assuming only the fundamental RP illumination profile. If an arbitrary cylindrically-symmetric phase map $\Phi$ is applied to the input beam before the focusing lens (sketched in Fig.~\ref{fig:spherical_field}, top), then there is no longer an analytical equation for the longitudinal field around the focus. However, using vector diffraction theory~\cite{stratton39,richards59,youngworth00} we can derive an integral description. Past work has studied the effects of aberrations in linearly polarized pulses~\cite{kempe93} and for radial polarization and tight focusing~\cite{biss04,singh08,april11,anita14,gaffar15,gaffar16}, but here we specifically focus on the longitudinal field, which was not emphasized in the past theory and is additionally significantly more difficult to measure experimentally.

The non-paraxial vector diffraction integral is based on the opening angle $\alpha$ and what is called the confocal parameter "$a$" that defines the level of focusing, as follows

\begin{align}
	\begin{split}
		\label{eq:E_z_arb}
		&E_z^{(\textrm{arb})}(z,t)=\sqrt{\frac{8P}{\pi\varepsilon_0 c}}\sqrt{\frac{k_0a}{2}}k_0e^{i(\omega_0 t+\Phi_0)}\\
		&\times\int_{0}^{\alpha_\textrm{max}}q(\alpha)\ell(\alpha)\sin^2(\alpha) e^{-i\omega_0z\cos(\alpha)/c}e^{i\Phi}\\
		&\qquad\qquad\times e^{-(t-z\cos(\alpha)/c+\Phi/\omega_0-F)^2/\tau^2}\,d\alpha,
	\end{split}
\end{align}

\noindent where $q(\alpha)=\textrm{sech}^2(\alpha/2)$ is an apodization function for a parabolic mirror and $\ell(\alpha)$ is the illumination profile. In this non-paraxial case the standard RPLB illumination is $\ell(\rho)=\rho e^{-\rho^2}$ where $\rho=r/w_i=\sqrt{2k_0a}\tan(\alpha/2)$. The aberration phase profile $\Phi(\rho)$ is left arbitrary for now. We can even consider an arbitrary pulse-front delay at the focusing optic $F(\rho)$ or a group-delay dispersion $\phi_2$, which will initially be set to zero but considered later on. The latter means $\tau^2=\tau_0^2+2i\phi_2$, i.e. a temporal chirp and increase in duration in the case of non-zero $\phi_2$. Those terms add no computational complexity since we know their expressions in time and have to do the integral over the input aperture anyway, and may eventually be further levers for optimization as seen previously~\cite{jolly19-1}.

There is an important nuance in this description, which is the inclusion of the phase terms also in the temporal envelope. This reflects the reality that any optic that deforms the phase must also have a temporal or chromatic effect, which is reflected in a minor pulse-front distortion at the input pupil proportional exactly to the phase. This includes also the main focusing optic, in addition to optic(s) that add aberration(s), for example a deformable mirror. The case that we show here assumes that the input beam size is independent of frequency, equivalent to the $g_0=1$ case when dealing with the focal phase~\cite{porras09,jolly21-1}, where we have shown that the effects are only significant for the shortest of pulses~\cite{jolly20-2}, but we still include the terms here for completeness.

In the paraxial regime $\alpha_\textrm{max}\ll1$ and we can make the approximations that $\sin(\alpha)\approx\alpha$, $\cos(\alpha)\approx1-\alpha^2/2$ (and therefore $q(\alpha)\approx1$), and $a\approx z_R$ (at the central frequency). We can also recognize that $\rho=r/w_i\approx\alpha\sqrt{k_0a/2}$, making the standard RPLB illumination much simpler. Although there exist more simplified solutions based on the error function in the case of purely spherical aberration, shown in appendix~\ref{sec:appendix_A}, they are still based on an integral. Therefore there is no practical reason to simplify to the paraxial case, nor necessarily for one specific aberration, so for the acceleration simulations we will use the full non-paraxial integral in Eq.~\ref{eq:E_z_arb}.

The phase map with purely spherical aberration can be written $\Phi=\sigma\rho^4$. We can see the numerical results of the integration in Fig.~\ref{fig:spherical_field}(a--b). As the spherical aberration $\sigma$ is decreased from zero to negative values in Fig.~\ref{fig:spherical_field}(a), the peak field $E_z$ is reduced and pushed to positive $z$, and has oscillations beyond the peak position. Note that the push towards positive $z$ depends on our chosen convention for the sign of the phase, and the fact that we model the spherical aberration with a pure quartic spatial phase---if it was modeled with a Zernike polynomial then the push away from $z=0$ would be almost completely removed. The characteristic length over which the field decays also becomes larger, which is in line with the well-known fact that spherical aberration can produce extended focal regions~\cite{sunQ05,smartsev19,oubrerie22,blum26}. However, since our application of electron acceleration depends on the peak field strength, we don't necessarily need a significantly extended focal region and will consider $|\sigma|$ values that are smaller than for that application. For the same values the phase along propagation is shown in Fig.~\ref{fig:spherical_field}(b), where the zero case is the well-known Gouy phase. With small values of $\sigma$ we also see oscillations in this phase, but at larger values the phase strongly exceeds the standard maximum of $\pi$. In fact, there is a singular transition where these phase oscillations change qualitatively occuring at $\sigma\approx-0.2795$, which we will discuss in more depth later. Although the paraxial model doesn't actually explicitly depend on $k_0a$ besides the scaling of the amplitude, extending to the non-paraxial case will begin to produce differences with the tightest of focusing (smaller $k_0a$). This is another reason to still use the non-paraxial integral in our simulations, given the sensitivity of electron trajectories to small changes in the field.

\section{Optimization of acceleration with spherical aberration}

We now turn to acceleration simulations using the integral representation for the field with spherical aberration. Still with electrons having no initial velocity, we repeat the optimization over the initial position $z(0)$ and CEP $\Phi_0$, and now additionally over the dimensionless spherical aberration parameter $\sigma$. We use the same numerical methods as for past works~\cite{jolly19-1,jolly20-2} and the results of Fig.~\ref{fig:background}, the relativistic Lorentz force and a fifth-order Adams-Bashforth algorithm for the forward finite difference solving.

\begin{figure*}[tb]
	\centering
	\includegraphics[width=177mm]{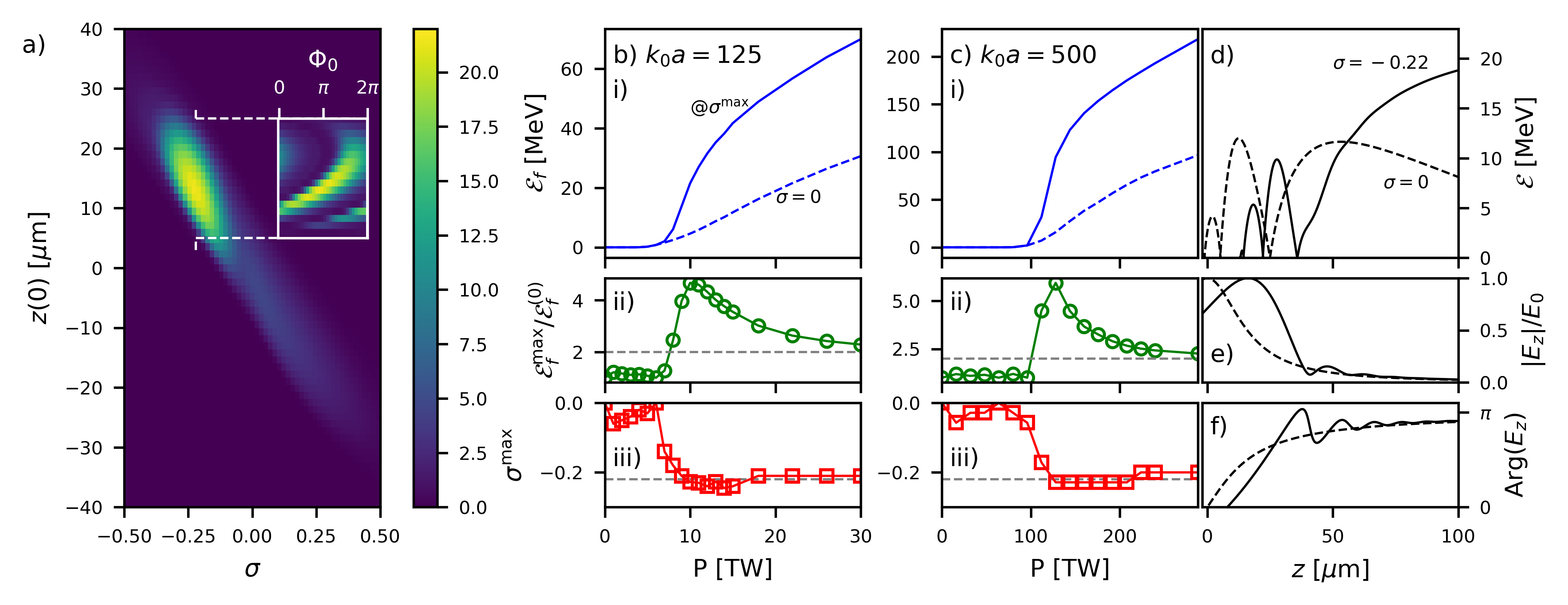}
	\caption{Optimization of the maximum electron energy with spherical aberration. The optimization space over $\{z(0),\sigma\}$ (a) for $\tau_0=10$\,fs, $k_0a=125$, and $P=10$\,TW showing significant improvement at negative spherical aberration $\sigma$ (each point is optimized over the CEP $\Phi_0$, where the inset shows that behavior at $\sigma=-0.22$). This is expanded to a large number of laser powers in (b) and for another focusing strength (c), showing the energy optimization over $\{\Phi_0,z(0),\sigma\}$ (i), the ratio of energy improvement with spherical aberration (ii), and finally at which value $\sigma^\textrm{max}$ that maximum occurs (iii). The electron energy during acceleration (d) is shown for the same case as in the inset of (a), together with the longitudinal field amplitude (e) and phase (f) for that specific case.}
	\label{fig:spherical_acceleration}  
\end{figure*}

As shown in Fig.~\ref{fig:spherical_acceleration}(a) at 10\,TW, with $\tau_0=10$\,fs and $k_0a=125$ (roughly a waist of 2\,$\mu$m), there is a significant increase of the maximum final electron energy around $\sigma=-0.2$. We choose these parameters because they are similar to previous works, the focusing level is not unrealistically tight, but that tight focusing allows for reasonably low laser powers, and finally the combination of short laser duration and tight focusing allows for shorter simulations. Importantly, the optimal aberration must have negative sign, where the interesting features are pushed towards positive $z$ in Fig.~\ref{fig:spherical_field} i.e. the direction of acceleration. The inset of Fig.~\ref{fig:spherical_acceleration}(a) shows the CEP behavior at the optimal $\sigma=-0.22$, where we see similar peaked behavior expected with significant acceleration as seen in Fig.~\ref{fig:background}(b).

Repeating the same optimization process at a large range of laser powers, we see in Fig.~\ref{fig:spherical_acceleration}(b.i) that the threshold power is not decreased, but above that threshold the spherical aberration robustly increases the maximum final kinetic energy of the test electron. We see in Fig.~\ref{fig:spherical_acceleration}(b.ii) that the improvement is the largest just above the threshold (at around 10\,TW) but still remains far above the threshold at around a factor of 2. Also, interestingly, the magnitude of the spherical aberration that produces the maximum final kinetic energy $\sigma^\textrm{max}$ stays constant for all powers above the threshold, as seen in Fig.~\ref{fig:spherical_acceleration}(b.iii).

Results for a looser focusing are shown in Fig.~\ref{fig:spherical_acceleration}(c) for the same $\tau_0=10$\,fs with $k_0a=500$ (roughly a waist of 4\,$\mu$m). We see as we expect a significantly higher threshold power of roughly 90\,TW (the scaling would predict a $\times16$ increase when compared to the previous case). With the same optimization over $\{\Phi_0,z(0),\sigma\}$ we see a similar behavior as with tighter focusing. The spherical aberration does not increase the threshold, but results in a robust improvement beyond the threshold as seen in Fig.~\ref{fig:spherical_acceleration}(c.i). The same analysis of the relative improvement due to spherical aberration in Fig.~\ref{fig:spherical_acceleration}(c.ii) shows that it is actually a larger difference with looser focusing. Still, as shown in Fig.~\ref{fig:spherical_acceleration}(c.iii), the magnitude of the optimal spherical aberration $\sigma^\textrm{max}$ is roughly the same at around $-0.22$. In addition we have found that with a tighter focusing of $k_0a=32$ (roughly a waist of 1\,$\mu$m), not shown here, the trend is the same---the spherical aberration allows for robust improvement once above the threshold, with roughly the same optimal $\sigma$ value, and with the relative effect smaller for tighter focusing.

The fact that the optimal spherical aberration is always of the same rough magnitude hints clearly at some sort of universal effect. We turn to looking at the specific electron trajectories to trace the source of the increase in the final electron energies. The trajectory without any spherical aberration shown in the dashed line of Fig.~\ref{fig:spherical_acceleration}(d) is an electron starting at $z(0)=-2\,\mu$m and the laser having a CEP of 5.7\,rad. (although the change with CEP is minor in this case), while the solid line trajectory with $\sigma=-0.22$ is an electron starting at $z(0)=13\,\mu$m and the laser having a CEP of 2.8\,rad.

One can notice immediately in both cases the electron is accelerating and decelerating as the laser goes through its focus at $z=0$ and overtakes the electron. Also for both cases the final acceleration occurs well after the laser focus, which is a forgotten reality of RP-VLA, where phase slippage and the alternating acceleration and deceleration means that net acceleration requires a decreasing field amplitude and a flattening phase with propagation. Without spherical aberration the electron reaches above 10\,MeV, but is still subsequently decelerated to the final energy below 5\,MeV. However, with spherical aberration, this deceleration no longer occurs to any significant extent and the electron is further accelerated to above 20\,MeV. The reason for this behavior with spherical aberration can only be deduced when looking in parallel at the accelerating field magnitude and phase in Fig.~\ref{fig:spherical_acceleration}(e--f). Although the field amplitude oscillates, it is the phase oscillations afforded by the spherical aberration that allow the electron to avoid the decelerating cycles and be continuously accelerated. The best accelerated electron shown in Fig.~\ref{fig:spherical_acceleration}(d) begins anew an accelerating cycle just before the first phase oscillation, so that it can most optimally be matched to subsequent phase oscillations.

The reason for the universal effect of spherical aberration is because these oscillations in the phase occur with the same values of spherical aberration regardless of the focusing strength. They do occur at earlier or later $z$, scaling by the value of $k_0a$, but they allow for optimal acceleration in exactly the same way. In fact, looking at the phase for a number of $\sigma$ values in Fig.~\ref{fig:spherical_field}(b) there is a singularity that occurs at -0.2795 such that the first phase oscillation disappears. Beyond that value there are further oscillations (not shown) that still allow for an improvement when compared to without spherical aberration, but that falls off very quickly around -0.4 as seen in Fig.~\ref{fig:spherical_acceleration}(a). So, this universal optimum in electron acceleration with spherical aberration is deeply related to the fundamental behavior of the (longitudinal) optical field with propagation, a behavior of which we have not encountered before and may have further impacts beyond this niche situation. When the laser power exceeds those levels studied here, it is now sensible that the improvement due to spherical aberration decreases, since the acceleration occurs over a shorter and shorter window of propagation. Note also that similar oscillations occur when the amplitude is strongly shaped~\cite{pelchat-voyer20}, meaning that it could be an alternate route to the same result or a point for future joint optimization with phase aberrations.

\begin{figure}[htb]
	\centering
	\includegraphics[width=86mm]{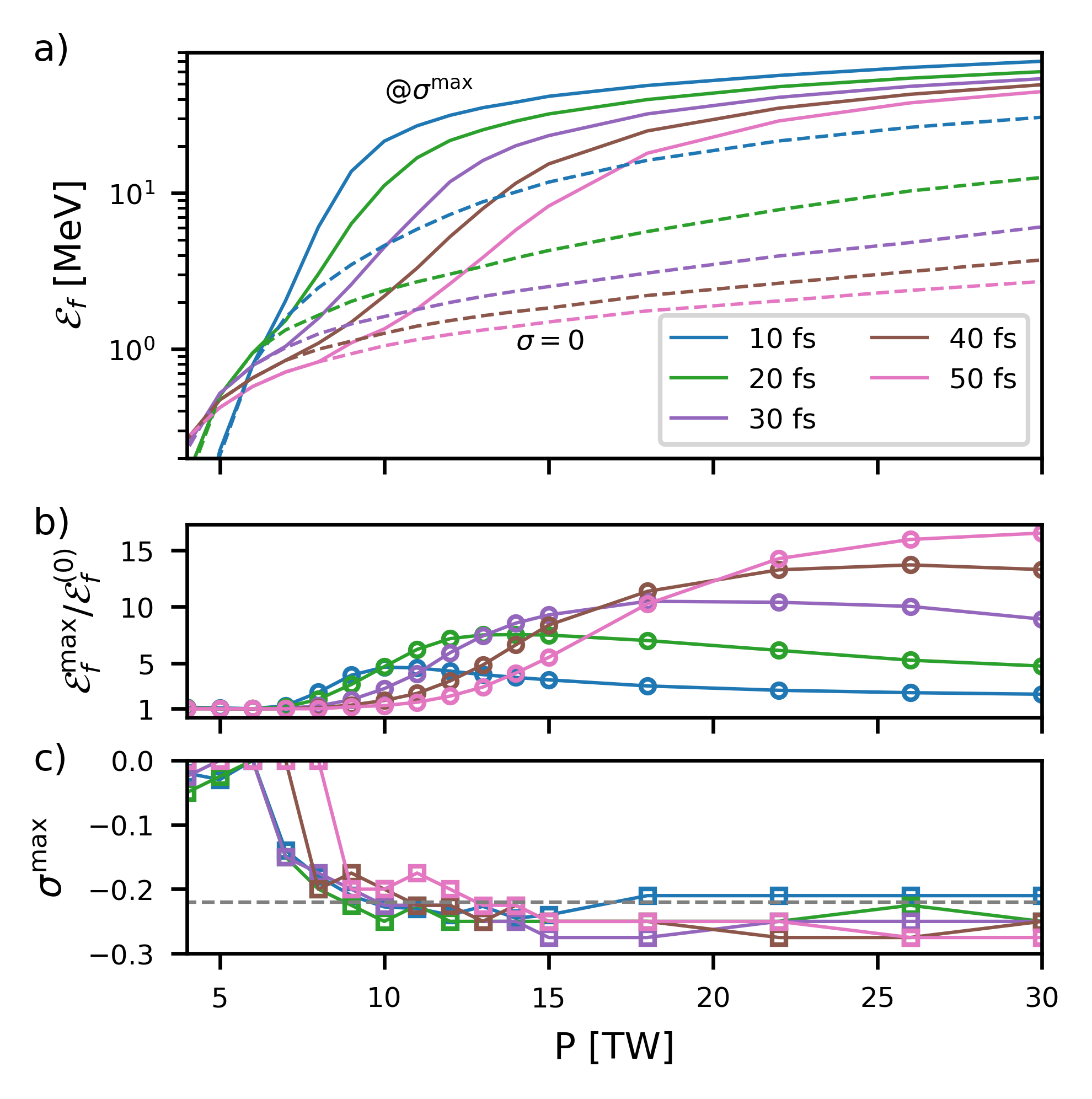}
	\caption{Behavior of the maximum electron energy (a) with spherical aberration (solid lines) for a large range of laser powers and pulse durations for a focusing level of $k_0a=125$. The relative improvement with spherical aberration (b) significantly increases with pulse duration, while the value of the best aberration is roughly the same (c).}
	\label{fig:spherical_acceleration_duration}
\end{figure}

Taking a step back, we look again at the acceleration with the focusing level $k_0a=125$ but this time with different pulse durations. These results for a range of powers can be seen in Fig.~\ref{fig:spherical_acceleration_duration}. Past work showed that, although near the threshold the optimal duration is not necessarily easy to predict, above the threshold the maximum electron energy increases with decreasing duration~\cite{wong10}. This can be intuitively understood as any optical cycles that are not taking part in the final accelerating phase are therefore necessarily decelerating the electron that is experiencing the entire laser pulse. So, for a fixed power, a shorter duration isolates only the accelerating half-cycle as much as possible. We reproduce these same results without any spherical aberration when optimizing over $\{\Phi_0,z(0)\}$ (dashed lines in Fig.~\ref{fig:spherical_acceleration_duration}(a)). We see a similar qualitative effect when also maximizing over the spherical aberration $\sigma$, seen in the solid lines of Fig.~\ref{fig:spherical_acceleration_duration}(a), but the magnitude of the effect of the pulse duration seems significantly weaker. Above threshold the longer pulse durations up to 50\,fs can accelerate to comparable energies as the shortest pulse duration of 10\,fs. 

When looking at the relative effect of the spherical aberration in Fig.~\ref{fig:spherical_acceleration_duration}(b), the longer pulse durations are significantly more affected. This can be understood as the fact that the phase oscillations with propagation of the aberrated pulse allow for the electron to experience the accelerating half cycle for much longer and in the end be decelerated much less or not at all by subsequent laser cycles, thereby reducing the advantage of shorter pulses. The improvement afforded by spherical aberration to a 50\,fs duration laser pulse is more than a factor of 15. As the pulse duration increases the maximum improvement occurs at increasing laser power. This may have a significant impact on the practicality of the RP-VLA mechanism, i.e. no longer requiring pulses of such a short duration to have efficient acceleration. This is especially true when considering that available pulse power from international laser facilities is actually the largest for "intermediate" pulse durations of 20--30\,fs, and some technologies even provide world-class power levels at longer durations up to 100--200\,fs. Going beyond the highest laser powers available, the Terawatt power levels investigated here are also much more common and easily attainable with standard 20--30\,fs duration titanium-Sapphire lasers, and becoming more attainable with longer duration ytterbium-based systems (with or without stand-alone pulse compression systems). However, CEP stabilization and control is still highly relevant, and is something that is less common and straightforward in intermediate pulse duration laser systems. As a final check we can see in Fig.~\ref{fig:spherical_acceleration_duration}(c) that for all durations tested the optimal spherical aberrations above the threshold is always of a similar magnitude, confirming once again the universal effect regardless of both the focusing level and pulse duration.

\section{Attempt to globally optimize with phase aberrations}

\begin{figure*}[tb]
	\centering
	\includegraphics[width=177mm]{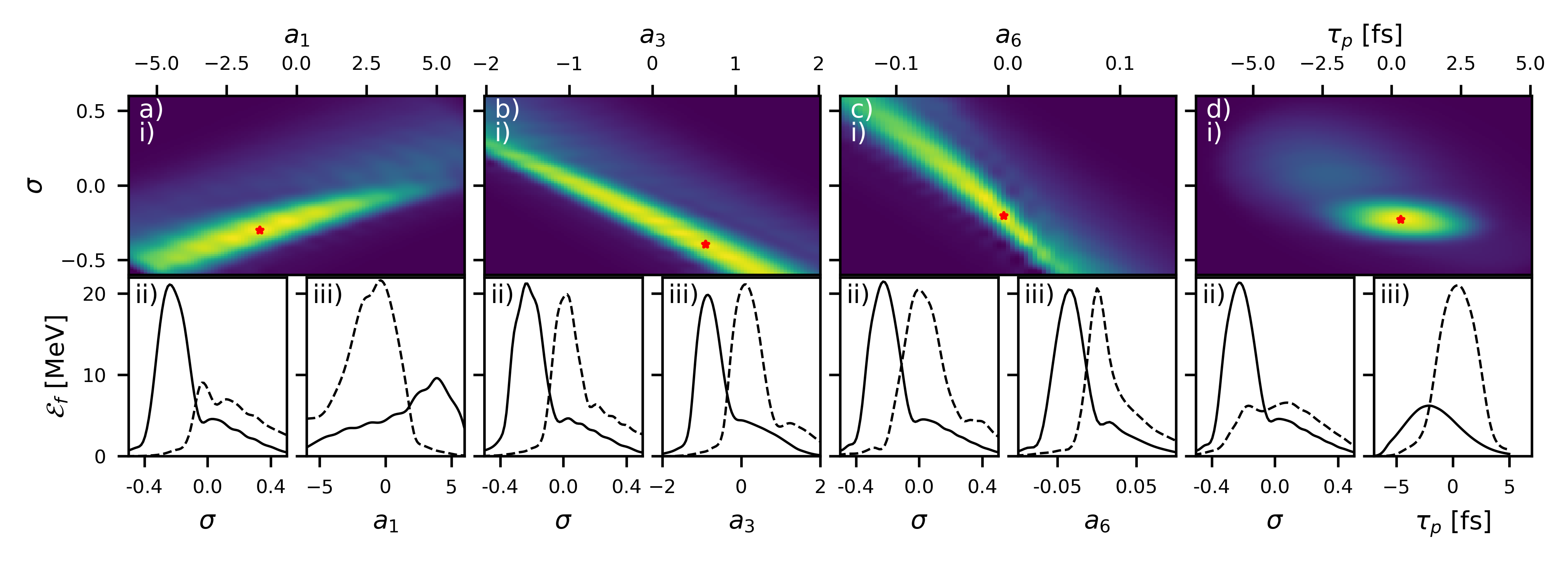}
	\caption{Optimization of electron energy with spherical aberration $\sigma$ combined with other aberration orders (each point is optimized over $\{z(0),\Phi_0\}$) for the axicon aberration $a_1$ (a), cubic $a_3$ (b), second-order spherical $a_6$ (c), and pulse-front curvature $\tau_p$ (d) with red stars denoting the global maximum. All solid lines are with the conjugate optimization parameter at zero. The dashed line in (a.ii) is with $a_1=4.5$, in (b.ii) it is with $a_3=-0.9$, in (c.ii) it is with $a_6=-0.036$, and in (d.ii) it is with $\tau_p=-2.5$\,fs. The dashed lines in all of the (iii) panels are with $\sigma=-0.25$. The laser power is 10\,TW, the duration is 10\,fs, and the focusing strength is $k_0a=125$, i.e. the same case as in Fig.~\ref{fig:spherical_acceleration}(a,d).}
	\label{fig:arbitrary_acceleration}
\end{figure*}

We discussed when introducing the model that, since we regardless use the non-paraxial and general integral to calculate the accelerating field, that any type of aberration can be modeled. We first considered spherical aberration as it is practical and intuitive, but in this section we will consider other arbitrary orders of phase aberrations and even pulse-front aberrations to attempt to optimize even further from the results in the previous section. We do not need to make any changes to the model to do that, so there is no increase in computational complexity, but indeed the number of free parameters increases significantly.

Returning to the integral description of the longitudinal field in Eq.~\ref{eq:E_z_arb}, the phase map $\Phi(\rho)$ can be written as a sum of weighted polynomials in $\rho$ such that $\Phi(\rho)=\sum a_n\rho^n$. This is an alternative to using the purely radial Zernike polynomials, that allows one to more freely make arbitrary phase maps, admittedly without considering whether they are practical or not. The term $a_1$ is a linear phase that has a discontinuity in its derivative at the origin, and is equivalent to the phase acquired from an axicon lens. The term $a_2$ is exactly the same as a standard lens and only shifts the interaction along $z$ so we will not consider it, and $a_3$ is a cubic phase front. Spherical aberration $\sigma=a_4$ and second-order spherical aberration can be related to $a_6$. Since we consider spherical aberration to be the base case, we will first perform brute-force optimization scans over $\sigma=a_4$ together with each of $\{a_1,a_3,a_6\}$ independently of each other.

The results of optimization over $\{z(0),\Phi_0,\sigma,a_{1,3,6}\}$ can be seen in Fig.~\ref{fig:arbitrary_acceleration}(a--c) for a single laser power of 10\,TW and a focusing level of $k_0a=125$, where each point on the maps (i) is the maximum over $\{z(0),\Phi_0\}$. The maxima that we observe occur at $\{\sigma,a_1\}=\{-0.3,-1.32\}$, $\{\sigma,a_3\}=\{-0.396,0.64\}$, and $\{\sigma,a_6\}=\{-0.204,-0.004\}$. We note a number of things---the global optima are either not far from the optimum with only spherical aberration or produce a maximum energy that is only marginally improved, or both; the optima with only $a_1$, $a_3$, or $a_6$ do not produce higher energies than with only spherical aberration; the optimization space of $\{\sigma,a_{1,3,6}\}$ is quite flat around the optimum, and extends along a region where there is some linear constraint between the two phase aberrations considered. The values of the maxima are summarized in Table~\ref{table1}. Another curious behavior is the reversal in the orientation of that linear constraint for the axicon phase $a_1$, which is due to the fact that it is the sole phase aberration that is of lower order than the fundamental focusing of the light $\propto\rho^2$.

In sum, these observations tell us that spherical aberration is the best aberration by itself for RP-VLA (but only slightly), and that each aberration by itself is doing roughly the same thing as shown in Fig.~\ref{fig:spherical_acceleration}(d--f). This is why in the optimization space it appears like the aberrations simply take the place of each other. A global optimum can be found that is some combination of the aberrations, but because they are all optimizing the electron acceleration in the same way, this global optimum doesn't give a significantly larger electron energy. The phase oscillations of Fig.~\ref{fig:spherical_acceleration}(f) are slightly changing in form to match the highly nonlinear process of RP-VLA, but not qualitatively changing. Therefore this specific case of optimizing RP-VLA can be seen as an inefficient optimization problem, since the fitness function (the final electron energy maximized over the initial position and CEP) extends among the many variables but is actually quite flat, meaning it is difficult to globally optimize but that optimization only slightly improves the fitness. The physical reason for this, which we will return to later, is that these plots are hiding the optimization space over $z(0)$, such that each different combination of aberrations that results in almost the same final energy is actually for a different $z(0)$.

We also investigate the joint optimization of the electron energy with spherical aberration and pulse-front curvature, i.e. where $F=\tau_p\rho^2$ in Eq.~\ref{eq:E_z_arb}, in Fig.~\ref{fig:arbitrary_acceleration}(d). A reminder of the improvement with initial pulse-front aberration is shown in appendix~\ref{sec:appendix_B}, using the specific algorithm of this work and in a larger range of parameters than in the past~\cite{jolly19-1}. To limit the dimensionality of the optimization space we will consider the pulse chirp still to be zero, although that can provide minor additional improvements. In this case there are two different types of aberrations, so we expect to see a qualitative difference in the optimization space when compared to joint optimization with two phase aberrations. We do indeed see in the map of Fig.~\ref{fig:arbitrary_acceleration}(d.i) two distinct peaks at the optimum for spherical aberration and the optimum for pulse-front curvature. However, the pulse-front curvature engenders a much smaller improvement itself compared to the spherical aberration, and the two effects seemingly do not have a hidden synergy. The global maximum occurs at $\{\sigma,\tau_p\}=\{-0.228,0.32$\,fs$\}$, i.e. very close to the maximum with purely spherical aberration and only larger by a small amount. This data point is also seen in the summary of Table~\ref{table1}.

\begin{table}[h]
\centering
\begin{tabular}{c|c|c|c|c|c|c}
	  & none & $a_1$ & $a_3$ & $a_6$ & $\tau_p$ \\
	  \hline
	none & 4.4 & 9.6 & 19.8 & 20.5 & 6.2 \\
	\hline
	$\sigma$ & 21.2 & 22 & 21.6 & 22.1 & 21.8
\end{tabular}
\caption{The final kinetic energy of electrons maximized over $\{z(0),\Phi_0\}$ in MeV for different discrete combinations of aberrations for 10\,TW and $k_0a=125$. The aberration values for these maxima are in the text.}
\label{table1}
\end{table}

The natural next question is if a much more complex set of aberrations and optimization in that much more highly dimensional space could produce much higher electron energies. Doing so would require much more sophisticated optimization algorithms to have a realistic computational time, which is becoming more present in science in general but also specifically in ultrafast light-matter interaction physics~\cite{miller26}. We approach this problem using the optimization space over all standard phase aberrations from $a_1$ through $a_8$ (except the bare curvature $a_2$) and the CEP $\Phi_0$ and initial electron position $z(0)$. We choose to not consider pulse-front aberrations since they are much more complicated to impart experimentally and likely have a smaller effect given our evidence in Appendix~\ref{sec:appendix_B} and Fig.~\ref{fig:arbitrary_acceleration}(d).

\begin{figure}[htb]
	\centering
	\includegraphics[width=86mm]{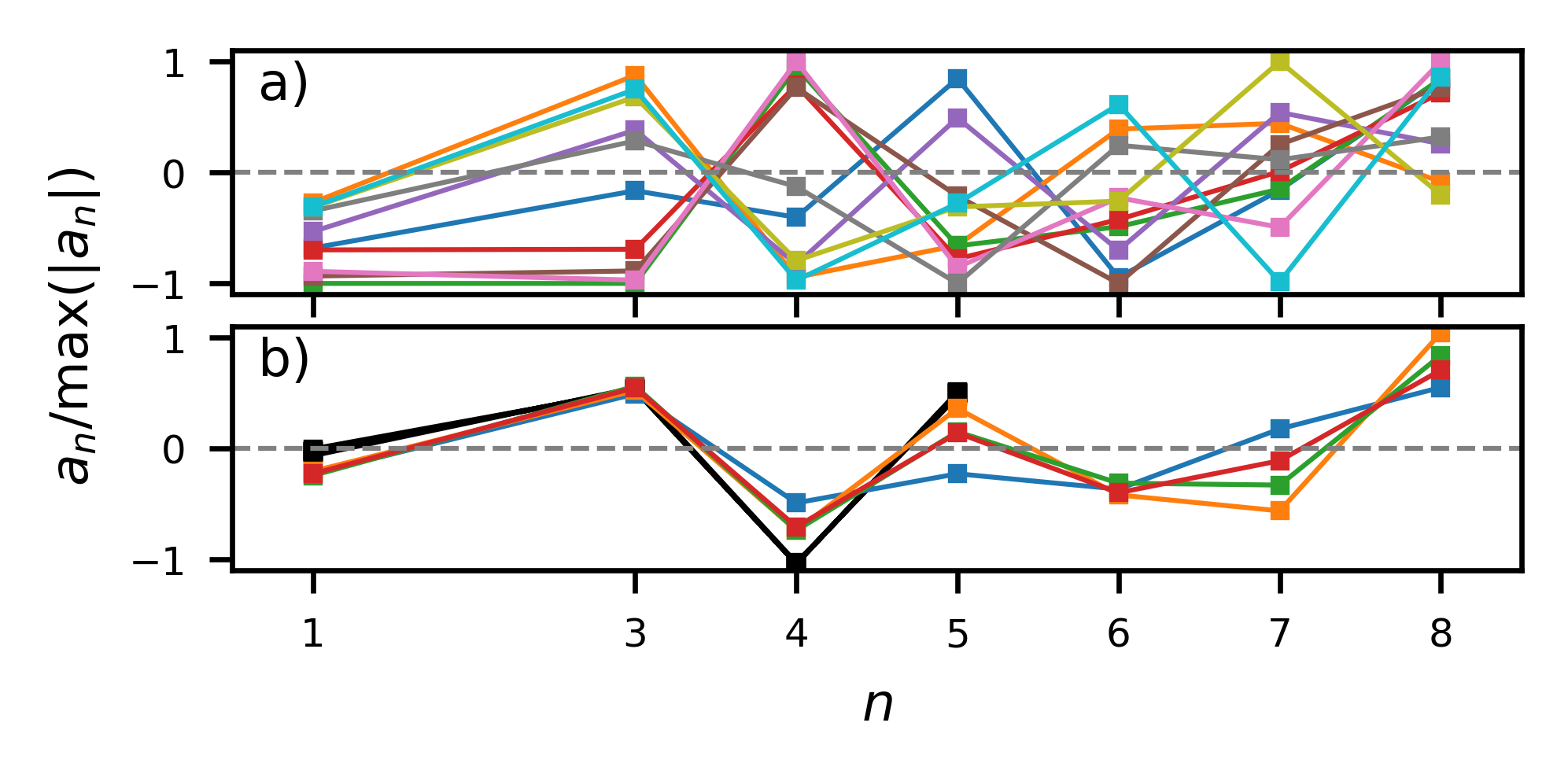}
	\caption{Resulting normalized best aberration parameters (a) for ten optimizations of the electron energy with 10\,TW and $k_0a=125$ using differential evolution while keeping $z(0)$ a free parameter, all resulting in $23.7\pm0.1$\,MeV. When $z(0)=0$ is kept fixed (b), the optimal aberrations are more closely packed while still reaching the same electron energy. With only 4 aberrations (black) the optimal aberrations are almost identical.}
	\label{fig:optimization}
\end{figure}

For a given set of aberrations we perform first a brute-force scan over $\Phi_0$ since we know \textit{a priori} that the dependence can be complicated and strongly peaked (see Fig.~\ref{fig:background}(b) and the inset of Fig.~\ref{fig:spherical_acceleration}(a)) and therefore difficult to optimize over. That result, the maximum electron kinetic energy optimized over the CEP $\Phi_0$, becomes our fitness function (or more precisely the negative of that number, since optimization algorithms are designed to minimize the fitness). With that fitness we perform differential evolution to optimize over the remaining eight dimensions---seven aberrations and the initial electron position. The results can be seen in Fig.~\ref{fig:optimization}(a) for 10\,TW and $k_0a=125$ (excluding the initial electron position since that is not necessarily a control parameter in eventual experiments, and many values would be sampled at once). For the shown results at every point 21 different CEP values are sampled, and the fitness is calculated 20,000-30,000 times depending on the speed of convergence, meaning each found set of optimal aberrations is based on around 500,000 simulated laser-electron interactions. After the initial differential evolution algorithm, which is inherently stochastic, we also perform a final deterministic minimization (referred to as "polishing"), in our case using the L-BFGS-B algorithm.

We note a few important details of this eight-dimensional optimization. First, we show ten different runs of the differential evolution algorithm that all find different sets of aberrations (see Fig.~\ref{fig:optimization}(a)) while still being considered successful. Secondly, in all ten cases the electron energy is optimized to be $23.7\pm0.1$\,MeV. So all cases are optimized further from the simple two-aberration cases shown before, and they are all somewhat equivalent meaning that either there is no global optimum and rather a large set of equivalent local optima, or the global optimum is very difficult to find. Importantly note that the "polishing" step has still not caused the results to cluster on one single global optimum, meaning that the optimal surface must have some structure. And finally, the improvement to 23.7\,MeV is only marginal compared to all previous optima found in Table~\ref{table1}. Note that the region where an aberration has an effect becomes smaller and smaller with the order of the aberration, as already seen in Fig.~\ref{fig:arbitrary_acceleration} and intuitive since they are based on the polynomial order of the spatial phase. Higher-order aberrations still play an important role but naturally take on smaller absolute magnitudes.

The above conclusions cement further our observation that the optimization space is somewhat "flat," although not perfectly so, and we now see that there are also near-optimal surfaces in different regions of the optimization space. Looking at the dependence of the optimal aberrations on each other, we see when comparing $a_1-a_3$ and $a_1-\sigma$ that there is a transition point at $a_1=-2$ where the optimal $a_3$ and $a_4=\sigma$ switch signs, i.e. the optimum is in a different region. However, not shown in Fig.~\ref{fig:optimization} is that all of those very different phase polynomials actually result in roughly the same evolution of the phase with propagation (only slightly different from that with purely spherical aberration) but shifted along $z$ and having different optimal $z(0)$. If we constrain $z(0)=0$ and do the same optimization, Fig.~\ref{fig:optimization}(b), then we see that with seven aberrations there are still multiple found optima---the problem is over-determined---but their differences are less significant than with $z(0)$ left free, and mostly in the higher-order aberrations. Upon reducing the number of aberrations to four the optima are always the same and the maximum electron energy is reduced to 22.3\,MeV. Physically we can now understand this in the context of an interesting feature of diffraction, that a large set of phase maps can result in almost identical propagation behavior, and that the different optima found may all be ways to create the lone global optimum of propagation behavior. Our interpretation of this optimization is that we only need a certain number of aberrations to optimize the process, and that if we are, for example, ionizing electrons at a given position then there may only be one best set of aberrations, otherwise there may be a large number of possible best aberrations.

\section{Discussion and conclusion}

In this work we have shown that phase aberrations on an ultrashort radially-polarized laser beam can significantly improve the acceleration in vacuum of electrons from rest from within the focus of said laser beam. The strong longitudinal field produced when tightly focusing such laser pulses allows for acceleration, and the phase aberrations produce phase oscillations upon propagation that allow for the acceleration process to continue advantageously.

We first build up our results by discussing the model, a combination of new physics in the calculation of the longitudinal field with aberrations and known physics in the relativistic Lorentz force. Then we apply the model to purely spherical aberration, where we see strong optimization of up to many times larger maximum electron energy, and study the behavior for different laser powers, focusing strengths, and pulse durations. Surprisingly, since the aberration allows for the acceleration process to build up over many laser cycles, we show that with the aberration applied there is a significantly decreased advantage to having a shorter driving pulse duration. Put another way, optimizing with the aberration is more efficient with longer pulse durations, and may even result in this process becoming accessible for more standard laser systems as long as they have CEP control.

Moving beyond only spherical aberration, we showed that the combination of aberrations can result in slightly higher electron energy, although with only marginal improvements and in a relatively trivial optimization space. Expanding to an eight-dimensional optimization space (including seven aberrations) and performing a stochastic optimization algorithm, we see that the optimization space is actually not at all trivial. We find many separate local optima. However, these local optima have once again only slightly improved maximum electron energies over the case of pure spherical aberration.

Although the RP-VLA mechanism is relatively niche, it stills shows promise especially for creating useful relativistic electron beams with very short duration---potentially based on the optical cycle time-scale rather than the laser pulse duration. Additionally, from a more fundamental point of view, RP-VLA is one of the few true ultrafast field-sensitive interactions together with high-harmonic generation and strong-field ionization, for example. Therefore in our present study we have uncovered a new behavior when adding phase aberrations, that links the transverse spatial extent of the input laser beam to the longitudinal coordinate where both the laser and accelerated electrons co-propagate, while also providing a motivation for the future applicability of RP-VLA. We considered mainly phase aberrations since they can be engendered in a relatively simple way with a deformable mirror, for example, and have no inherent losses. We observed robust increase in accelerated electron energy above the threshold, but it is a very interesting problem whether the threshold for significant net acceleration could also be reached by lower laser power. In the future we could also expand the optimization space even further to include spatial amplitude or spatio-spectral/spatio-temporal phase and/or amplitude, consider aberrations that break cylindrical symmetry, and use alternative fitness functions like electron beam energy-spread, divergence, or pulse duration. This would allow us to both further study this interesting field-sensitive effect and in parallel interrogate its ceiling of performance for future applications.

\section*{Acknowledgments}

The author acknowledges François Fillion-Gourdeau for discussions on optimization algorithms.

\section*{Funding}

The author acknowledges funding from Les Fonds de la Recherche Scientifique (F.R.S.-FNRS).

\appendix

\section{Fields with spherical aberration in terms of the error function}
\label{sec:appendix_A}

Here we will restrict the phase aberration to be equivalent to spherical aberration such that $\Phi=\sigma\rho^4$, and the pulse-front aberration is purely quadratic pulse-front curvature $F=\tau_p\rho^2$. Applying all of the paraxial approximations, the longitudinal field on-axis is then

\begin{align}
	\begin{split}
		\label{eq:E_z_spherical}
		&E_z^{(\textrm{sph})}(z,t)=\sqrt{\frac{8P}{\pi\varepsilon_0 c}}\frac{2e^{i(\omega_0(t-z/c)+\Phi_0)}}{z_R}\\
		&\times\int_{0}^{\infty}\rho^3 e^{\rho^2(iz/z_R-1)+i\sigma\rho^4}\\
		&\qquad\times e^{-(t-z/c+[z/\omega_0z_R-\tau_p]\rho^2+\sigma\rho^4/\omega_0)^2/\tau^2}\,d\rho.
	\end{split}
\end{align}

\noindent The integration range is set to go to $+\infty$, meaning that we will not including effects related to a limited aperture or clipping of the illumination profile, which is also realistic in the paraxial or slightly non=paraxial cases.

If we assume that the pulse-front delay due to the spherical aberration is negligible, i.e. $\sigma\ll\omega_0\tau_0$, then we can ignore that term in the temporal envelope, and group terms of the paraxial integral Eq.~\ref{eq:E_z_spherical} in terms of $\rho^2$ and $\rho^4$. Making the variable change to $u=\rho^2$ we can write the resultant integral in a general form as follows

\begin{align}
E_z^{(\textrm{sph})}(z,t)&=E_0\int_{0}^{\infty}u e^{-p^2 u^2 - qu}\,du,\\
p^2&=\left(\frac{z}{\omega_0\tau z_R}-\frac{\tau_p}{\tau}\right)^2-i\sigma,\\
q&=1-\frac{iz}{z_R}+2t^\prime\left(\frac{z}{\omega_0\tau^2 z_R}-\frac{\tau_p}{\tau^2}\right),
\end{align}

\noindent where $E_0$ includes the electric field amplitude, oscillations, and temporal envelope and we define $t^\prime=t-z/c$. Note that if the beam is very long ($\omega_0\tau_0\rightarrow\infty$) and $\{\sigma,\phi_2,\tau_p\}$ are all zero the integral is approximately equal to $1/q^2=1/(1-iz/z_R)^2$, which gives us Eq.~\ref{eq:E_z_an}.

There is a known solution to the integral using the error function, solution 3.462.5 in Ref.~\cite{gradshteyn14}

\begin{align}
	\begin{split}
	&E_z^{(\textrm{sph})}(z,t)=\\
	&\qquad\frac{E_0}{2p^2}\left(1-\frac{q\sqrt{\pi}}{2p}e^{q^2/4p^2}\left[1-\textrm{erf}\left(\frac{q}{2p}\right)\right]\right).
	\end{split}
\end{align}

\noindent As stressed in the main text, although these solutions are interesting, the error function is still an integral and therefore this description does not necessarily improve our computation of the field for accelerating the electron.

In fact, a similar method can be used to solve for the transverse electric field of a linearly-polarized pulse-beam with Gaussian illumination in the same configuration based on solution 3.322.2 in Ref.~\cite{gradshteyn14}

\begin{align}
	\begin{split}
	E_\perp^{(\textrm{LP-sph})}(z,t)&=E_0^{\textrm{LP}}\int_{0}^{\infty}e^{-p^2 u^2 - qu}\,du,\\
	&=\frac{\sqrt{\pi}E_0^{\textrm{LP}}}{2p}e^{q^2/4p^2}\left[1-\textrm{erf}\left(\frac{q}{2p}\right)\right].
	\end{split}
\end{align}

\noindent In this case of linear polarization, this on-axis solution can more easily be extended to an off-axis solution as well, since we are operating in the paraxial approximation and so there is just the one scalar component.

It is unclear how novel these solutions are, since they present pseudo-analytical descriptions of ultrafast pulse-beams with spherical aberration and even including the focal phase for the $g_0=1$ case (when the input beam size is independent of frequency) which seems novel even if $\sigma=0$. The full context and implications thereof are beyond the scope of this work, but we note in conclusion that there are indeed numerous efficient algorithms for calculating the error function, so there may be eventual computational speed-up in specific cases or even in general if that avenue is pursued.

\section{Acceleration with longitudinal chromatism}
\label{sec:appendix_B}

\begin{figure}[tb]
	\centering
	\includegraphics[width=86mm]{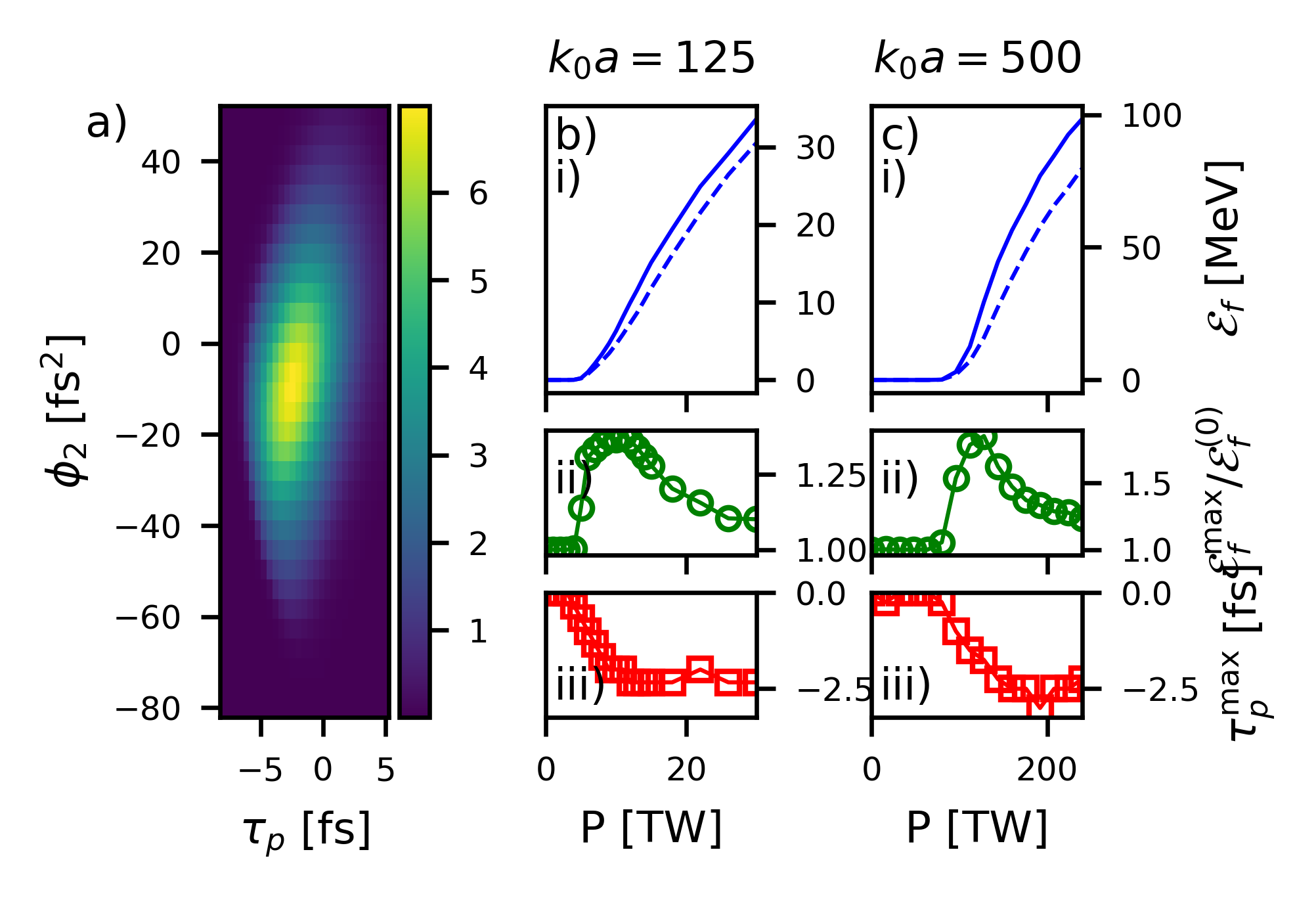}
	\caption{Electron energy with pulse-front curvature on the input beam. The optimization space over $\{\phi_2,\tau_p\}$ (a) for $\tau_0=10$\,fs and $k_0a=125$ showing improvement at non-zero value for both optimization variables (each point is optimized over $\{z(0),\Phi_0\}$). This is expanded to a large number of laser powers without chirp in (b) and for another focusing strength (c), showing the energy optimization (i), the ratio of energy improvement with $\tau_p$ (ii), and finally at which value $\tau_p^\textrm{max}$ that the maximum occurs (iii).}
	\label{fig:LC_acceleration}
\end{figure}

To add context to the main text, we detail here the effect of longitudinal chromatism at the focus on the RP-VLA acceleration mechanism. This concept was first reported in past work~\cite{jolly19-1}, but in that case the modeling was done via a temporal Fourier transform at the focus rather than a diffraction integral as done in this work. Therefore the first role of this appendix is to confirm that the main behavior stays the same, both with the different model and with slightly different laser parameters. These results are summarized in Fig.~\ref{fig:LC_acceleration}.

When using the diffraction integral we are modeling effects on the input illumination rather than on the focused beam, so the longitudinal chromatism modeled in the previous work is modeled here by the quadratic pulse-front curvature parameter $\tau_p$ with $F=\tau_p\rho^2$, as already explained in the main text. The main surprising result of acceleration in this case is that first, it can be improved, and second, that it can be further improved with pulse chirp $\phi_2$ (Fig.~\ref{fig:LC_acceleration}(a)). Both of these modifications to the driving pulse decrease the peak accelerating field but still result in a higher final electron energy. Pulse chirp by itself will always decrease the accelerated electron energy, but when combined with pulse-front curvature can provide slight additional optimization. These results agree with the previous work, with the main difference that here we consider tighter focusing (and therefore lower powers), and the relevant assumptions on the spatio-spectral amplitude are slightly different. Therefore the magnitude of the effect is only slightly changed.

When focusing only on the longitudinal chromatism (pulse-front curvature), we take the opportunity also to look at a larger range of laser powers, and two focusing levels (and therefore beam waists) in Fig.~\ref{fig:LC_acceleration}(b--c). We see in Fig.~\ref{fig:LC_acceleration}(b.i) and (c.i) that, just as with spherical aberration, the final electron energy increases but only above the threshold, and the threshold itself is not significantly affected. We note that when comparing Fig.~\ref{fig:LC_acceleration}(b.ii) and (c.ii) that the strength of the effect increases with looser focusing, also similar to with spherical aberration.

\end{document}